\documentclass[amsmath,aps,showpacs,a4paper,10pt]{revtex4}

 \usepackage{epsf}
 \usepackage{graphicx}    

 \usepackage{longtable}   

\begin{document}

 \newcommand{\be}[1]{\begin{equation}\label{#1}}
 \newcommand{\ee}{\end{equation}}
 \newcommand{\bea}{\begin{eqnarray}}
 \newcommand{\eea}{\end{eqnarray}}
 \def\disp{\displaystyle}

 \def\gsim{ \lower .75ex \hbox{$\sim$} \llap{\raise .27ex \hbox{$>$}} }
 \def\lsim{ \lower .75ex \hbox{$\sim$} \llap{\raise .27ex \hbox{$<$}} }

 \begin{titlepage}

 \begin{flushright}
 arXiv:1004.4951
 \end{flushright}

 \title{\Large \bf Observational Constraints on Cosmological
 Models with the~Updated Long Gamma-Ray Bursts}

 \author{Hao~Wei\,}
 \email[\,email address:\ ]{haowei@bit.edu.cn}
 \affiliation{Department of Physics, Beijing Institute
 of Technology, Beijing 100081, China}

 \begin{abstract}\vspace{1cm}
 \centerline{\bf ABSTRACT}\vspace{2mm}
 In the present work, by the help of the newly released Union2
 compilation which consists of 557 Type Ia supernovae (SNIa),
 we calibrate 109 long Gamma-Ray Bursts (GRBs) with the
 well-known Amati relation, using the cosmology-independent
 calibration method proposed by Liang {\it et al.}. We have
 obtained 59 calibrated high-redshift GRBs which can be used
 to constrain cosmological models without the circularity
 problem (we call them ``Hymnium'' GRBs sample
 for convenience). Then, we consider the joint constraints on
 7 cosmological models from the latest observational data,
 namely, the combination of 557 Union2 SNIa dataset, 59
 calibrated Hymnium GRBs dataset (obtained in this work), the
 shift parameter $R$ from the WMAP 7-year data, and the
 distance parameter $A$ of the measurement of the baryon
 acoustic oscillation (BAO) peak in the distribution of SDSS
 luminous red galaxies. We also briefly consider the comparison
 of these 7 cosmological models.
 \end{abstract}


 \maketitle

 \end{titlepage}

 \renewcommand{\baselinestretch}{1.0}


\section{Introduction}\label{sec1}

Since the discovery of current accelerated expansion of our
 universe~\cite{r1}, Type Ia supernovae (SNIa) have been
 considered to be a powerful probe to study this mysterious
 phenomenon. However, SNIa are plagued with extinction from
 the interstellar medium, and hence the current maximum
 redshift of SNIa is only about $z\simeq 1.755$. On the other
 hand, the redshift of the last scattering surface of cosmic
 microwave background (CMB) is about $z\simeq 1090$. There is
 a wide ``desert'' between the redshifts of SNIa and CMB. So,
 the observations at intermediate redshift are important to
 distinguish cosmological models.

Recently, Gamma-Ray Bursts (GRBs) have been proposed to be a
 complementary probe to SNIa (see e.g.~\cite{r2} and references
 therein). So far, GRBs are the most intense explosions
 observed in our universe. Their high energy photons in the
 gamma-ray band are almost immune to dust extinction, in
 contrast to supernovae. Up to now, there are many GRBs observed at
 $0.1<z\leq 8.1$, whereas the maximum redshift of GRBs is
 expected to be $10$ or even larger~\cite{r3}. Therefore, GRBs
 are considered to be a hopeful probe to fill the ``desert''
 between the redshifts of SNIa and CMB. We refer
 to e.g.~\cite{r2,r4,r5,r6} for some comprehensive reviews on
 the so-called GRB cosmology. As is well known,  there is a
 circularity problem in the direct use of GRBs~\cite{r7},
 mainly due to the lack of a set of low-redshift GRBs at
 $z<0.1$ which are cosmology-independent. To alleviate the
 circularity problem, some statistical methods have been
 proposed, such as the scatter method~\cite{r8}, the
 luminosity distance method~\cite{r8}, and the Bayesian
 method~\cite{r9}. Other methods trying to avoid the
 circularity problem have been proposed in e.g.~\cite{r10,r11}.
 Recently, a new idea of the distance ladder to calibrate GRBs
 in a completely cosmology-independent manner has been proposed
 in~\cite{r12,r13} independently. Similar to the case of
 calibrating SNIa as secondary standard candles by using Cepheid
 variables which are primary standard candles, we can also
 calibrate GRBs as standard candles with a large amount of
 SNIa. And then, the calibrated GRBs can be used to constrain
 cosmological models without the circularity problem. We refer
 to e.g.~\cite{r12,r14,r15} for some relevant works.

It is worth noting that in the literature most relevant works
 mainly used the 69 GRBs compiled in~\cite{r2} or the 70 GRBs
 compiled in~\cite{r16}. As of the end of 2009, there are 109
 long GRBs with measured redshift and spectral peak energy
 compiled in~\cite{r17}. The number of usable GRBs is
 significantly increased~\cite{r60}. Among these 109 GRBs, the
 data of 70 GRBs are taken from~\cite{r16}; the data of 25 GRBs
 are taken from~\cite{r18}; the data of remaining 14 GRBs
 (090516, 090618, 090715B, 090812, 090902B, 090926, 090926B,
 091003, 091018, 091020, 091024, 091029, 091127, 091208B) are
 provided by L.~Amati in private communication~\cite{r19}. On
 the other hand, the cosmology-independent method to calibrate
 GRBs proposed in~\cite{r12} needs a large amount of SNIa at
 $z<1.4$. Thus, the SNIa dataset plays an important role in the
 calibration of GRBs. Previously, the relevant works
 (e.g.~\cite{r12,r14,r15}) used the 192 Davis SNIa dataset or
 the 307 Union SNIa dataset to calibrate GRBs. Very recently,
 the Supernova Cosmology Project (SCP) collaboration released
 their Union2 compilation which consists of 557
 SNIa~\cite{r20}. The Union2 compilation is the largest
 published and spectroscopically confirmed SNIa sample to date.
 The number of usable SNIa is also significantly increased.
 Obviously, a larger SNIa dataset could bring a better
 interpolation for the cosmology-independent method to
 calibrate GRBs proposed in~\cite{r12}. Motivated by both the
 significant improvements in GRBs and SNIa, in the present
 work, we update the calibration of GRBs with the well-known
 Amati relation, using the cosmology-independent method
 proposed in~\cite{r12}, and obtain 59 calibrated
 high-redshift GRBs which can be used to constrain cosmological
 models without the circularity problem (we call them ``Hymnium''
 GRBs sample for convenience).

On the other hand, very recently the WMAP Collaboration also
 released their 7-year CMB data (WMAP7) in~\cite{r21}.
 Therefore, it is natural to consider the joint constraints on
 cosmological models from these latest observational data,
 namely, the combination of 557 Union2 SNIa dataset~\cite{r20},
 59 calibrated Hymnium GRBs dataset (obtained in this work),
 the shift parameter $R$ from the WMAP7 data~\cite{r21}, and
 the distance parameter $A$ of the measurement of the BAO peak
 in the distribution of SDSS luminous red galaxies~\cite{r22,r23}.
 In this work, we consider 7 cosmological models and obtain the
 observational constraints on them. Notice that we assume the
 universe to be spatially flat throughout this work.

This paper is organized as followings. In Sec.~\ref{sec2}, by
 the help of the newly released Union2 compilation which
 consists of 557 SNIa, we calibrate the 109 GRBs with Amati
 relation, using the cosmology-independent method proposed
 in~\cite{r12}. In Sec.~\ref{sec3}, we briefly introduce the
 methodology to constrain cosmological models with the combined
 latest observational data. Then, we obtain the corresponding
 constraints on 7 cosmological models in Sec.~\ref{sec4}.
 Further, we consider the comparison of these 7 models in
 Sec.~\ref{sec5}. Finally, we give the brief conclusions in
 Sec.~\ref{sec6}.


\section{Updating the calibration of GRBs with Union2 SNIa data}\label{sec2}

In this section, as mentioned above, we are going to calibrate
 the 109 GRBs~\cite{r17} (and~\cite{r16,r18,r19}) by using the
 557 Union2 SNIa data~\cite{r20}. In fact, we are closely
 following the cosmology-independent calibration method used
 in~\cite{r12,r15}. Notice that in the present work, we
 calibrate GRBs only with the Amati relation, so that we
 can use a larger GRBs dataset for single GRB luminosity
 relation. As in~\cite{r12,r15}, we choose $z=1.4$ to be the
 divide line to separate GRBs into low- and high-redshift
 groups. In the 109 GRBs compiled by Amati~\cite{r17}
 (and~\cite{r16,r18,r19}), there are 50 GRBs at $z<1.4$ and 59
 GRBs at $z>1.4$. The maximum redshift of these 109 GRBs is
 $z=8.1$ for GRB~090423.

Several years ago, Amati {\it et al.} found the well-known
 $E_{\rm p,i}-E_{\rm iso}$ correlation in GRBs as
 $E_{\rm p,i}=K\times E_{\rm iso}^m$ by using 12 GRBs with
 known redshifts~\cite{r24} (note that this correlation has
 been firstly discovered by Lloyd~{\it et~al.}~\cite{r62}
 independently), where $E_{\rm p,i}=E_{\rm p,obs}\times (1+z)$
 is the cosmological rest-frame spectral peak energy; the
 isotropic-equivalent radiated energy is given by
 \be{eq1}
 E_{\rm iso}=4\pi\,d_L^2\,S_{\rm bolo}\,(1+z)^{-1},
 \ee
 in which $S_{\rm bolo}$ is the bolometric fluence of gamma
 rays in the GRB at redshift $z$, and $d_L$ is the luminosity
 distance of the GRB. Later, Amati {\it et al.} have updated it
 in~\cite{r25,r16,r18,r17}. Up to now, some theoretical
 interpretations have been proposed for the Amati
 relation~\cite{r4}. It might be geometrical effects due to
 the jet viewing angle with respect to a ring-shaped emission
 region~\cite{r26}, or with respect to a multiple sub-jet model
 structure~\cite{r27}. An alternative explanation of the Amati
 relation is related to the dissipative mechanism responsible
 for the prompt emission~\cite{r28}. However, we alert the
 readers to the fact that there is a considerable debate as to
 whether the Amati relation is an intrinsic effect or the
 result of detection biases (or a combination of these
 two)~\cite{r63}. We refer to e.g.~\cite{r61,r64} for some
 notable works which discuss the Amati relation in particular.
 They noted that many outliers are not included in the studies
 of Amati and Schaefer, and argued that this relation might
 simply be wrong, or its inferred scatter might be too low or
 different from the observed scatter~\cite{r63}. Anyway, let
 us go ahead. For convenience, similar to~\cite{r2}, we can
 rewrite the Amati relation as
 \be{eq2}
 \log\frac{E_{\rm iso}}{\rm erg}=\lambda+
 b\,\log\frac{E_{\rm p,i}}{\rm \,300\,keV\,}\,,
 \ee
 where ``$\log$'' indicates the logarithm to base $10$, whereas
 $\lambda$ and $b$ are constants to be determined. In the
 literature, the Amati relation was calibrated with
 $E_{\rm iso}$ computed by {\em assuming} a $\Lambda$CDM
 cosmology with specified model parameters. As mentioned
 above, this is cosmology-dependent and the circularity
 problem follows. Here, we instead use the method proposed
 in~\cite{r12} to calibrate the Amati relation in a
 cosmology-independent manner. We refer to the original
 paper~\cite{r12} for more details.

\begin{center}
 \begin{figure}[htbp]
 \centering
 \includegraphics[width=1.0\textwidth]{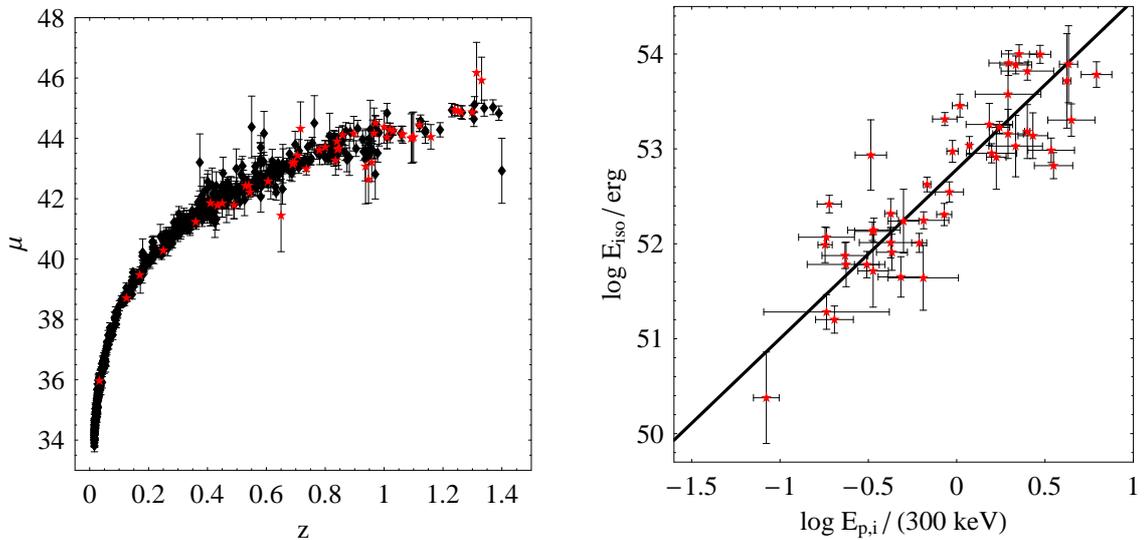}
 \caption{\label{fig1}
 Left panel: The Hubble diagram of 557 SNIa (black diamonds)
 and 50 low-redshift GRBs (red stars) whose distance moduli
 are derived by using cubic interpolation. Right panel: 50 GRBs
 data (red stars) in the
 $\log E_{\rm p,i}\,/({\rm 300\,keV})-\log E_{\rm iso}\,/{\rm erg}$
 plane. The best-fit calibration line is also plotted. See the
 text for details.}
 \end{figure}
 \end{center}

Firstly, we derive the distance moduli for the 50 low-redshift
 ($z<1.4$) GRBs of~\cite{r17} (and~\cite{r16,r18,r19})
 by using cubic interpolation from the 557 Union2 SNIa compiled
 in~\cite{r20}. We present the interpolated distance moduli
 $\mu$ of these 50 GRBs in the left panel of Fig.~\ref{fig1}.
 The corresponding error bars are also plotted. In
 Table~\ref{tab1}, we give the numerical data of these 50 GRBs.
 As in~\cite{r12}, when the cubic interpolation is used, the
 error of the distance modulus $\mu$ for the GRB at redshift
 $z$ can be calculated by
 $$\sigma_\mu=\left(\sum\limits_{i=1}^4 A_i^2
 \epsilon_{\mu,i}^2\right)^{1/2},~~~~~~~{\rm where}~~~~~
 A_i\equiv\left.\prod\limits_{j\not=i}\left(z_j-z\right)
 \right/\prod\limits_{j\not=i}\left(z_j-z_i\right),$$
 in which $j$ runs from 1 to 4 but $j\not=i$; on the other
 hand, $\epsilon_{\mu,i}$ are the errors of the nearby SNIa
 whose redshifts are $z_i$. Then, using the well-known relation
 \be{eq3}
 \mu=5\log\frac{d_L}{\rm \,Mpc}+25\,,
 \ee
 one can convert distance modulus $\mu$ into luminosity
 distance $d_L$ (in units of Mpc). From Eq.~(\ref{eq1}) with
 the corresponding $S_{\rm bolo}$ given in~\cite{r16,r18,r19},
 we can derive $E_{\rm iso}$ for these 50 GRBs at $z<1.4$.
 We present them in the right panel of Fig.~\ref{fig1}, whereas
 $E_{\rm p,i}$ for these 50 GRBs at $z<1.4$ are taken
 from~\cite{r16,r18,r19}. Also, we present the errors for these
 50 GRBs at $z<1.4$, by simply using the error propagation. From
 Fig.~\ref{fig1}, one can clearly see that the intrinsic
 scatter is dominating over the measurement errors. Therefore,
 as in~\cite{r2,r12}, the bisector of the two ordinary least
 squares~\cite{r29} will be used. Following the procedure of
 the bisector of the two ordinary least squares described
 in~\cite{r29}, we find the best fit to be
 \be{eq4}
 b=1.7828~~~~~~~{\rm and}~~~~~~~\lambda=52.7838\,,
 \ee
 with $1\sigma$ uncertainties
 \be{eq5}
 \sigma_b=0.0072~~~~~~~{\rm and}~~~~~~~\sigma_\lambda=0.0041\,.
 \ee
 The best-fit calibration line Eq.~(\ref{eq2}) with $b$ and
 $\lambda$ in Eq.~(\ref{eq4}) is also plotted in the right
 panel of Fig.~\ref{fig1}. From Eq.~(\ref{eq5}), it is easy to
 see that the calibration in this work is better than the one
 in~\cite{r15}.


\vspace{-3mm} 

\begin{center}
 \begin{figure}[htbp]
 \centering
 \includegraphics[width=0.45\textwidth]{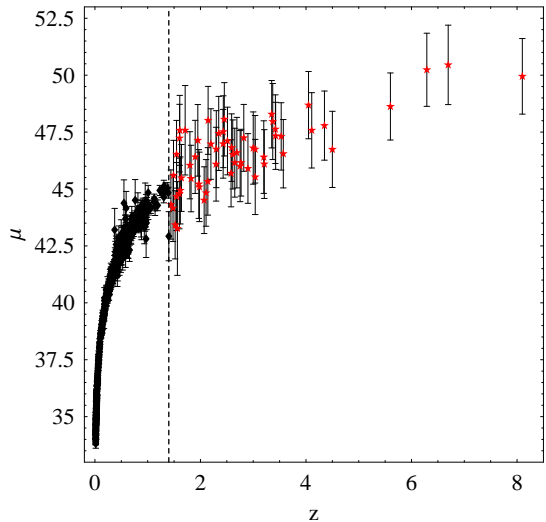}
 \caption{\label{fig2}
 The Hubble diagram of 557 Union2 SNIa (black diamonds)
 and 59 high-redshift Hymnium GRBs (red stars) whose distance moduli
 are derived by using the calibrated Amati relation. The dashed
 line indicates $z=1.4$. See the text for details.}
 \end{figure}
 \end{center}

Next, we extend the calibrated Amati relation to high-redshift,
 namely $z>1.4$. Since $E_{\rm p,i}$ for the 59 GRBs at $z>1.4$
 have been given in~\cite{r16,r18,r19}, we can derive $E_{\rm iso}$
 from the calibrated Amati relation Eq.~(\ref{eq2}) with $b$ and
 $\lambda$ in Eq.~(\ref{eq4}). Then, we derive the distance moduli
 $\mu$ for these 59 GRBs at $z>1.4$ using Eqs.~(\ref{eq1})
 and~(\ref{eq3}) while their $S_{\rm bolo}$ can be taken
 from~\cite{r16,r18,r19}. On the other hand, the propagated
 uncertainties are given by~\cite{r2}
 \be{eq6}
 \sigma_\mu=\left[\left(\frac{5}{2}\sigma_{\log E_{\rm iso}}\right)^2
 +\left(\frac{5}{2\ln 10}\,\frac{\sigma_{S_{\rm bolo}}}{S_{\rm bolo}}
 \right)^2\right]^{1/2},
 \ee
 where
 \be{eq7}
 \sigma_{\log E_{\rm iso}}^2=\sigma_\lambda^2+\left(\sigma_b
 \log\frac{E_{\rm p,i}}{\rm \,300\,keV\,}\right)^2+\left(
 \frac{b}{\ln 10}\,\frac{\sigma_{E_{\rm p,i}}}{E_{\rm p,i}}\right)^2
 +\sigma_{E_{\rm iso,sys}}^2,
 \ee
 in which $\sigma_{E_{\rm iso,sys}}$ is the systematic error and it
 accounts the extra scatter of the luminosity relation. As
 in~\cite{r2}, by requiring the $\chi^2/dof$ of the 50 points at
 $z<1.4$ in the
 $\log E_{\rm p,i}\,/({\rm 300\,keV})-\log E_{\rm iso}\,/{\rm erg}$
 plane about the best-fit calibration line to be unity, we find that
 \be{eq8}
 \sigma_{E_{\rm iso,sys}}^2=0.1526.
 \ee
 Note that in principle $\sigma_{E_{\rm iso,sys}}^2$ is a free
 parameter. However, if we permit it to vary with cosmology,
 like in e.g.~\cite{r65}, there might be room for the circularity
 problem. Even one does not care this problem, the constraints
 on cosmological models become loose, mainly due to the fact
 that the number of free parameters has been increased. On the
 other hand, in the cosmology-independent calibration method
 proposed in~\cite{r12}, we have not used any cosmology when we
 calibrate GRBs at $z<1.4$, unlike in~\cite{r65} they calibrate
 all GRBs without using SNIa at low-redshift. Mainly due to the
 nature of the calibration method proposed in~\cite{r12}, we
 have no freedom to determine $\sigma_{E_{\rm iso,sys}}^2$
 by cosmology, and hence we must use the method in~\cite{r2} to
 fix it by requiring $\chi^2/dof=1$. We admit that this
 prevents us to learn the systematics dominating the Amati
 relation. Nevertheless, we plot the derived distance moduli
 $\mu$ with $1\sigma$ uncertainties for these 59 GRBs at
 $z>1.4$ in Fig.~\ref{fig2}. We also present the numerical
 data of these 59 GRBs in Table~\ref{tab2}. For convenience,
 we call them ``Hymnium'' GRBs sample. It is worth noting that
 they are obtained in a completely cosmology-independent
 manner, and hence can be used to constrain cosmological models
 without the circularity problem.


 \begingroup
 \renewcommand{\baselinestretch}{0.6}
 \squeezetable
 \begin{table}[ptbh]
 \begin{center}
 \begin{tabular}{llccc}
 \hline\hline \ GRB & ~~~~~$z$ & ~~~~$S_{\rm bolo}~(10^{-5}~\rm erg\, cm^{-2}$)
 & ~~~~~$E_{\rm p,i}~(\rm keV)$ & \ \ $\mu$ \\[1mm] \hline
 \ 060218	 & ~~~~ 0.0331	 & ~~~~ 2.20$\,\pm\,$0.10	 & ~~~~ 4.9$\,\pm\,$0.3	 & ~~~~ 35.97$\,\pm\,$0.22 \ \\
 \ 060614	 & ~~~~ 0.125	 & ~~~~ 5.90$\,\pm\,$2.40	 & ~~~~ 55$\,\pm\,$45	 & ~~~~ 38.71$\,\pm\,$0.11 \ \\
 \ 030329	 & ~~~~ 0.17	 & ~~~~ 21.50$\,\pm\,$3.80	 & ~~~~ 100$\,\pm\,$23	 & ~~~~ 39.47$\,\pm\,$0.15 \ \\
 \ 020903	 & ~~~~ 0.25	 & ~~~~ 0.016$\,\pm\,$0.004	 & ~~~~ 3.37$\,\pm\,$1.79	 & ~~~~ 40.29$\,\pm\,$0.13 \ \\
 \ 011121	 & ~~~~ 0.36	 & ~~~~ 24.30$\,\pm\,$6.70	 & ~~~~ 1060$\,\pm\,$265	 & ~~~~ 41.23$\,\pm\,$0.17 \ \\
 \ 020819B	 & ~~~~ 0.41	 & ~~~~ 1.60$\,\pm\,$0.40	 & ~~~~ 70$\,\pm\,$21	 & ~~~~ 41.86$\,\pm\,$0.21 \ \\
 \ 990712	 & ~~~~ 0.434	 & ~~~~ 1.40$\,\pm\,$0.30	 & ~~~~ 93$\,\pm\,$15	 & ~~~~ 41.78$\,\pm\,$0.26 \ \\
 \ 010921	 & ~~~~ 0.45	 & ~~~~ 1.80$\,\pm\,$0.20	 & ~~~~ 129$\,\pm\,$26	 & ~~~~ 41.85$\,\pm\,$0.46 \ \\
 \ 091127	 & ~~~~ 0.49	 & ~~~~ 2.34$\,\pm\,$0.28	 & ~~~~ 54$\,\pm\,$5	 & ~~~~ 41.79$\,\pm\,$0.45 \ \\
 \ 081007	 & ~~~~ 0.5295	 & ~~~~ 0.22$\,\pm\,$0.041	 & ~~~~ 61$\,\pm\,$15	 & ~~~~ 42.43$\,\pm\,$0.29 \ \\
 \ 090618	 & ~~~~ 0.54	 & ~~~~ 28.087$\,\pm\,$3.37	 & ~~~~ 257$\,\pm\,$41	 & ~~~~ 42.44$\,\pm\,$0.11 \ \\
 \ 090424	 & ~~~~ 0.544	 & ~~~~ 5.90$\,\pm\,$1.15	 & ~~~~ 273$\,\pm\,$50	 & ~~~~ 42.21$\,\pm\,$0.16 \ \\
 \ 050525A	 & ~~~~ 0.606	 & ~~~~ 2.60$\,\pm\,$0.50	 & ~~~~ 127$\,\pm\,$10	 & ~~~~ 42.58$\,\pm\,$0.33 \ \\
 \ 050416A	 & ~~~~ 0.65	 & ~~~~ 0.087$\,\pm\,$0.009	 & ~~~~ 25.1$\,\pm\,$4.2	 & ~~~~ 41.44$\,\pm\,$1.20 \ \\
 \ 080916	 & ~~~~ 0.689	 & ~~~~ 0.79$\,\pm\,$0.079	 & ~~~~ 184$\,\pm\,$18	 & ~~~~ 43.16$\,\pm\,$0.23 \ \\
 \ 020405	 & ~~~~ 0.69	 & ~~~~ 8.40$\,\pm\,$0.70	 & ~~~~ 354$\,\pm\,$10	 & ~~~~ 43.15$\,\pm\,$0.22 \ \\
 \ 970228	 & ~~~~ 0.695	 & ~~~~ 1.30$\,\pm\,$0.10	 & ~~~~ 195$\,\pm\,$64	 & ~~~~ 43.21$\,\pm\,$0.21 \ \\
 \ 991208	 & ~~~~ 0.706	 & ~~~~ 17.20$\,\pm\,$1.40	 & ~~~~ 313$\,\pm\,$31	 & ~~~~ 43.43$\,\pm\,$0.29 \ \\
 \ 041006	 & ~~~~ 0.716	 & ~~~~ 2.30$\,\pm\,$0.60	 & ~~~~ 98$\,\pm\,$20	 & ~~~~ 44.32$\,\pm\,$0.88 \ \\
 \ 090328	 & ~~~~ 0.736	 & ~~~~ 8.93$\,\pm\,$2.061	 & ~~~~ 1028$\,\pm\,$312	 & ~~~~ 42.99$\,\pm\,$0.20 \ \\
 \ 030528	 & ~~~~ 0.78	 & ~~~~ 1.40$\,\pm\,$0.20	 & ~~~~ 57$\,\pm\,$9	 & ~~~~ 43.61$\,\pm\,$0.17 \ \\
 \ 051022	 & ~~~~ 0.8	 & ~~~~ 32.60$\,\pm\,$3.10	 & ~~~~ 754$\,\pm\,$258	 & ~~~~ 43.71$\,\pm\,$0.22 \ \\
 \ 970508	 & ~~~~ 0.835	 & ~~~~ 0.34$\,\pm\,$0.07	 & ~~~~ 145$\,\pm\,$43	 & ~~~~ 43.26$\,\pm\,$0.48 \ \\
 \ 060814	 & ~~~~ 0.84	 & ~~~~ 3.80$\,\pm\,$0.40	 & ~~~~ 473$\,\pm\,$155	 & ~~~~ 43.90$\,\pm\,$0.23 \ \\
 \ 990705	 & ~~~~ 0.842	 & ~~~~ 9.80$\,\pm\,$1.40	 & ~~~~ 459$\,\pm\,$139	 & ~~~~ 43.63$\,\pm\,$0.53 \ \\
 \ 000210	 & ~~~~ 0.846	 & ~~~~ 8$\,\pm\,$0.90	 & ~~~~ 753$\,\pm\,$26	 & ~~~~ 43.67$\,\pm\,$0.70 \ \\
 \ 040924	 & ~~~~ 0.859	 & ~~~~ 0.49$\,\pm\,$0.04	 & ~~~~ 102$\,\pm\,$35	 & ~~~~ 44.11$\,\pm\,$0.30 \ \\
 \ 091003	 & ~~~~ 0.8969	 & ~~~~ 4.75$\,\pm\,$0.79	 & ~~~~ 810$\,\pm\,$157	 & ~~~~ 44.16$\,\pm\,$0.57 \ \\
 \ 080319B	 & ~~~~ 0.937	 & ~~~~ 49.70$\,\pm\,$3.80	 & ~~~~ 1261$\,\pm\,$65	 & ~~~~ 43.07$\,\pm\,$1.24 \ \\
 \ 071010B	 & ~~~~ 0.947	 & ~~~~ 0.74$\,\pm\,$0.37	 & ~~~~ 101$\,\pm\,$20	 & ~~~~ 42.64$\,\pm\,$0.79 \ \\
 \ 970828	 & ~~~~ 0.958	 & ~~~~ 12.30$\,\pm\,$1.40	 & ~~~~ 586$\,\pm\,$117	 & ~~~~ 43.20$\,\pm\,$0.64 \ \\
 \ 980703	 & ~~~~ 0.966	 & ~~~~ 2.90$\,\pm\,$0.30	 & ~~~~ 503$\,\pm\,$64	 & ~~~~ 44.17$\,\pm\,$0.84 \ \\
 \ 091018	 & ~~~~ 0.971	 & ~~~~ 0.30$\,\pm\,$0.03	 & ~~~~ 55$\,\pm\,$20	 & ~~~~ 44.51$\,\pm\,$0.23 \ \\
 \ 980326	 & ~~~~ 1	 & ~~~~ 0.18$\,\pm\,$0.04	 & ~~~~ 71$\,\pm\,$36	 & ~~~~ 44.38$\,\pm\,$0.53 \ \\
 \ 021211	 & ~~~~ 1.01	 & ~~~~ 0.42$\,\pm\,$0.05	 & ~~~~ 127$\,\pm\,$52	 & ~~~~ 44.04$\,\pm\,$0.38 \ \\
 \ 991216	 & ~~~~ 1.02	 & ~~~~ 24.80$\,\pm\,$2.50	 & ~~~~ 648$\,\pm\,$134	 & ~~~~ 44.29$\,\pm\,$0.21 \ \\
 \ 080411	 & ~~~~ 1.03	 & ~~~~ 5.70$\,\pm\,$0.30	 & ~~~~ 524$\,\pm\,$70	 & ~~~~ 44.25$\,\pm\,$0.14 \ \\
 \ 000911	 & ~~~~ 1.06	 & ~~~~ 23$\,\pm\,$4.70	 & ~~~~ 1856$\,\pm\,$371	 & ~~~~ 44.14$\,\pm\,$0.25 \ \\
 \ 091208B	 & ~~~~ 1.063	 & ~~~~ 0.79$\,\pm\,$0.056	 & ~~~~ 255$\,\pm\,$25	 & ~~~~ 44.12$\,\pm\,$0.28 \ \\
 \ 091024	 & ~~~~ 1.092	 & ~~~~ 16.57$\,\pm\,$1.60	 & ~~~~ 586$\,\pm\,$251	 & ~~~~ 44.00$\,\pm\,$0.82 \ \\
 \ 980613	 & ~~~~ 1.096	 & ~~~~ 0.19$\,\pm\,$0.03	 & ~~~~ 194$\,\pm\,$89	 & ~~~~ 44.02$\,\pm\,$0.84 \ \\
 \ 080413B	 & ~~~~ 1.1	 & ~~~~ 0.73$\,\pm\,$0.092	 & ~~~~ 150$\,\pm\,$30	 & ~~~~ 44.05$\,\pm\,$0.83 \ \\
 \ 000418	 & ~~~~ 1.12	 & ~~~~ 2.80$\,\pm\,$0.50	 & ~~~~ 284$\,\pm\,$21	 & ~~~~ 44.44$\,\pm\,$0.21 \ \\
 \ 061126	 & ~~~~ 1.1588	 & ~~~~ 8.70$\,\pm\,$1.00	 & ~~~~ 1337$\,\pm\,$410	 & ~~~~ 44.05$\,\pm\,$0.41 \ \\
 \ 090926B	 & ~~~~ 1.24	 & ~~~~ 0.83$\,\pm\,$0.042	 & ~~~~ 204$\,\pm\,$10	 & ~~~~ 44.94$\,\pm\,$0.19 \ \\
 \ 020813	 & ~~~~ 1.25	 & ~~~~ 16.30$\,\pm\,$4.10	 & ~~~~ 590$\,\pm\,$151	 & ~~~~ 44.92$\,\pm\,$0.18 \ \\
 \ 061007	 & ~~~~ 1.261	 & ~~~~ 21.10$\,\pm\,$2.10	 & ~~~~ 890$\,\pm\,$124	 & ~~~~ 44.87$\,\pm\,$0.21 \ \\
 \ 990506	 & ~~~~ 1.3	 & ~~~~ 21.70$\,\pm\,$2.20	 & ~~~~ 677$\,\pm\,$156	 & ~~~~ 44.87$\,\pm\,$0.22 \ \\
 \ 061121	 & ~~~~ 1.314	 & ~~~~ 5.10$\,\pm\,$0.60	 & ~~~~ 1289$\,\pm\,$153	 & ~~~~ 46.18$\,\pm\,$1.01 \ \\
 \ 071117	 & ~~~~ 1.331	 & ~~~~ 0.89$\,\pm\,$0.21	 & ~~~~ 647$\,\pm\,$226	 & ~~~~ 45.92$\,\pm\,$0.77 \ \\[1mm]
 \hline\hline
 \end{tabular}
 \end{center}
 \caption{\label{tab1} The numerical data of 50 low-redshift GRBs at
 $z<1.4$. The first 4 columns are taken from~\cite{r16,r18,r19},
 whereas the last column is derived by using cubic interpolation
 from the 557 Union2 SNIa. These 50 low-redshift GRBs can be used to
 calibrate the Amati relation. See the text for details.}
 \end{table}
 \endgroup



\vspace{-11mm} 


\section{Methodology to constrain cosmological models}\label{sec3}

Very recently, as mentioned in Sec.~\ref{sec1}, the SCP
 collaboration released their Union2 dataset which consists of
 557 SNIa~\cite{r20}, whereas the WMAP Collaboration also
 released their 7-year CMB data (WMAP7) in~\cite{r21}. On the
 other hand, as of the end of 2009, there are 109 long GRBs
 with measured redshift and spectral peak energy~\cite{r17}
 (and~\cite{r16,r18,r19}). The number of usable GRBs is
 significantly increased. Motivated by these significant
 updates in the observations of SNIa, CMB and GRBs, it is
 natural to consider the joint constraints on cosmological
 models with the latest observational data. Here, we use the
 combination of 557 Union2 SNIa dataset~\cite{r20}, 59
 calibrated Hymnium GRBs dataset (obtained in this work, and
 whose numerical data are given in Table~\ref{tab2}), the shift
 parameter $R$ from the WMAP7 data~\cite{r21}, and the distance
 parameter $A$ of the measurement of the BAO peak in the
 distribution of SDSS luminous red galaxies~\cite{r22,r23}.

The data points of the 557 Union2 SNIa compiled in~\cite{r20}
 and the 59 Hymnium GRBs dataset in Table~\ref{tab2} of this
 paper are given in terms of the distance modulus $\mu_{obs}(z_i)$.
 On the other hand, the theoretical distance modulus is defined as
 \be{eq9}
 \mu_{th}(z_i)\equiv 5\log_{10}D_L(z_i)+\mu_0\,,
 \ee
 where $\mu_0\equiv 42.38-5\log_{10}h$ and $h$ is the Hubble
 constant $H_0$ in units of $100~{\rm km/s/Mpc}$, whereas
 \be{eq10}
 D_L(z)=(1+z)\int_0^z \frac{d\tilde{z}}{E(\tilde{z};{\bf p})}\,,
 \ee
 in which $E\equiv H/H_0$, and ${\bf p}$ denotes the model
 parameters. Correspondingly, the $\chi^2$ from the 557 Union2
 SNIa and the 59 Hymnium GRBs is given by
 \be{eq11}
 \chi^2_{\mu}({\bf p})=\sum\limits_{i}\frac{\left[
 \mu_{obs}(z_i)-\mu_{th}(z_i)\right]^2}{\sigma^2(z_i)}\,,
 \ee
 where $\sigma$ is the corresponding $1\sigma$ error. The parameter
 $\mu_0$ is a nuisance parameter but it is independent of the data
 points. One can perform an uniform marginalization over $\mu_0$.
 However, there is an alternative way. Following~\cite{r30,r31}, the
 minimization with respect to $\mu_0$ can be made by expanding the
 $\chi^2_{\mu}$ of Eq.~(\ref{eq11}) with respect to $\mu_0$ as
 \be{eq12}
 \chi^2_{\mu}({\bf p})=\tilde{A}-2\mu_0\tilde{B}+\mu_0^2\tilde{C}\,,
 \ee
 where
 $$\tilde{A}({\bf p})=\sum\limits_{i}\frac{\left[\mu_{obs}(z_i)
 -\mu_{th}(z_i;\mu_0=0,{\bf p})\right]^2}
 {\sigma_{\mu_{obs}}^2(z_i)}\,,$$
 $$\tilde{B}({\bf p})=\sum\limits_{i}\frac{\mu_{obs}(z_i)
 -\mu_{th}(z_i;\mu_0=0,{\bf p})}{\sigma_{\mu_{obs}}^2(z_i)}\,,
 ~~~~~~~~~~
 \tilde{C}=\sum\limits_{i}\frac{1}{\sigma_{\mu_{obs}}^2(z_i)}\,.$$
 Eq.~(\ref{eq12}) has a minimum for
 $\mu_0=\tilde{B}/\tilde{C}$ at
 \be{eq13}
 \tilde{\chi}^2_{\mu}({\bf p})=
 \tilde{A}({\bf p})-\frac{\tilde{B}({\bf p})^2}{\tilde{C}}\,.
 \ee
 Since $\chi^2_{\mu,\,min}=\tilde{\chi}^2_{\mu,\,min}$
 obviously, we can instead minimize $\tilde{\chi}^2_{\mu}$
 which is independent of $\mu_0$. In fact, by this method,
 we have just marginalized over $\mu_0$ analytically, and
 then simplified our computing.

There are some other observational data relevant to this work,
 such as the observations of CMB anisotropy~\cite{r21} and
 large-scale structure (LSS)~\cite{r22,r23}. However, using
 the full data of CMB and LSS to perform a global fitting
 consumes a large amount of time and power. As an alternative,
 one can instead use the shift parameter $R$ from the CMB, and
 the distance parameter $A$ of the measurement of the baryon
 acoustic oscillation (BAO) peak in the distribution of SDSS
 luminous red galaxies. In the literature, the shift parameter
 $R$ and the distance parameter $A$ have been used extensively.
 It is argued that they are model-independent~\cite{r32}, while
 $R$ and $A$ contain the main information of the observations
 of CMB and BAO, respectively. As is well known, the shift
 parameter $R$ of the CMB is defined by~\cite{r32,r33}
 \be{eq14}
 R\equiv\Omega_{m0}^{1/2}\int_0^{z_\ast}
 \frac{d\tilde{z}}{E(\tilde{z})}\,,
 \ee
 where $\Omega_{m0}$ is the present fractional density of
 pressureless matter; the redshift of recombination $z_\ast=1091.3$
 which has been updated in the WMAP7 data~\cite{r21}. The shift
 parameter $R$ relates the angular diameter distance to
 the last scattering surface, the comoving size of the sound
 horizon at $z_\ast$ and the angular scale of the first
 acoustic peak in CMB power spectrum of temperature
 fluctuations~\cite{r32,r33}. The value of $R$ has been updated
 to $1.725\pm 0.018$ from the WMAP7 data~\cite{r21}. On the
 other hand, the distance parameter $A$ of the measurement of
 the BAO peak in the distribution of SDSS luminous red
 galaxies~\cite{r22} is given by
 \be{eq15}
 A\equiv\Omega_{m0}^{1/2}E(z_b)^{-1/3}\left[\frac{1}{z_b}
 \int_0^{z_b}\frac{d\tilde{z}}{E(\tilde{z})}\right]^{2/3},
 \ee
 where $z_b=0.35$. In~\cite{r23}, the value of $A$ has been
 determined to be $0.469\,(n_s/0.98)^{-0.35}\pm 0.017$. Here
 the scalar spectral index $n_s$ is taken to be $0.963$, which
 has been updated from the WMAP7 data~\cite{r21}. So, the total
 $\chi^2$ is given by
 \be{eq16}
 \chi^2=\tilde{\chi}^2_{\mu}+\chi^2_{CMB}+\chi^2_{BAO}\,,
 \ee
 where $\tilde{\chi}^2_{\mu}$ is given in Eq.~(\ref{eq13}),
 $\chi^2_{CMB}=(R-R_{obs})^2/\sigma_R^2$ and
 $\chi^2_{BAO}=(A-A_{obs})^2/\sigma_A^2$. The best-fit model
 parameters are determined by minimizing the total $\chi^2$.
 As in~\cite{r34,r66}, the $68\%$ confidence level is
 determined by $\Delta\chi^2\equiv\chi^2-\chi^2_{min}\leq 1.0$,
 $2.3$ and $3.53$ for $n_p=1$, $2$ and $3$, respectively, where
 $n_p$ is the number of free model parameters. On the other
 hand, the $95\%$ confidence level is determined by
 $\Delta\chi^2\equiv\chi^2-\chi^2_{min}\leq 4.0$, $6.17$ and
 $8.02$ for $n_p=1$, $2$ and $3$, respectively.


 \begingroup
 \renewcommand{\baselinestretch}{0.6}
 \squeezetable
 \begin{table}[ptbh]
 \begin{center}
 \begin{tabular}{llccc}
 \hline \hline \ GRB & ~~~~~$z$ & ~~~~$S_{\rm bolo}~(10^{-5}~\rm erg\, cm^{-2}$)
 & ~~~~~$E_{\rm p,i}~(\rm keV)$ & \ \ $\mu$ \\[1mm] \hline
 \ 050318	 & ~~~~ 1.44	 & ~~~~ 0.42$\,\pm\,$0.03	 & ~~~~ 115$\,\pm\,$25	 & ~~~~ 44.32$\,\pm\,$1.52 \ \\
 \ 010222	 & ~~~~ 1.48	 & ~~~~ 14.60$\,\pm\,$1.50	 & ~~~~ 766$\,\pm\,$30	 & ~~~~ 44.15$\,\pm\,$1.46 \ \\
 \ 060418	 & ~~~~ 1.489	 & ~~~~ 2.30$\,\pm\,$0.50	 & ~~~~ 572$\,\pm\,$143	 & ~~~~ 45.60$\,\pm\,$1.54 \ \\
 \ 030328	 & ~~~~ 1.52	 & ~~~~ 6.40$\,\pm\,$0.60	 & ~~~~ 328$\,\pm\,$55	 & ~~~~ 43.43$\,\pm\,$1.50 \ \\
 \ 070125	 & ~~~~ 1.547	 & ~~~~ 13.30$\,\pm\,$1.30	 & ~~~~ 934$\,\pm\,$148	 & ~~~~ 44.67$\,\pm\,$1.49 \ \\
 \ 090102	 & ~~~~ 1.547	 & ~~~~ 3.48$\,\pm\,$0.63	 & ~~~~ 1149$\,\pm\,$166	 & ~~~~ 46.53$\,\pm\,$1.49 \ \\
 \ 040912	 & ~~~~ 1.563	 & ~~~~ 0.21$\,\pm\,$0.06	 & ~~~~ 44$\,\pm\,$33	 & ~~~~ 43.27$\,\pm\,$2.06 \ \\
 \ 990123	 & ~~~~ 1.6	 & ~~~~ 35.80$\,\pm\,$5.80	 & ~~~~ 1724$\,\pm\,$466	 & ~~~~ 44.80$\,\pm\,$1.55 \ \\
 \ 071003	 & ~~~~ 1.604	 & ~~~~ 5.32$\,\pm\,$0.59	 & ~~~~ 2077$\,\pm\,$286	 & ~~~~ 47.23$\,\pm\,$1.48 \ \\
 \ 090418	 & ~~~~ 1.608	 & ~~~~ 2.35$\,\pm\,$0.59	 & ~~~~ 1567$\,\pm\,$384	 & ~~~~ 47.58$\,\pm\,$1.54 \ \\
 \ 990510	 & ~~~~ 1.619	 & ~~~~ 2.60$\,\pm\,$0.40	 & ~~~~ 423$\,\pm\,$42	 & ~~~~ 44.94$\,\pm\,$1.47 \ \\
 \ 080605	 & ~~~~ 1.6398	 & ~~~~ 3.40$\,\pm\,$0.28	 & ~~~~ 650$\,\pm\,$55	 & ~~~~ 45.49$\,\pm\,$1.47 \ \\
 \ 091020	 & ~~~~ 1.71	 & ~~~~ 0.11$\,\pm\,$0.034	 & ~~~~ 280$\,\pm\,$190	 & ~~~~ 47.58$\,\pm\,$1.96 \ \\
 \ 080514B	 & ~~~~ 1.8	 & ~~~~ 2.027$\,\pm\,$0.48	 & ~~~~ 627$\,\pm\,$65	 & ~~~~ 46.04$\,\pm\,$1.47 \ \\
 \ 090902B	 & ~~~~ 1.822	 & ~~~~ 32.38$\,\pm\,$1.01	 & ~~~~ 2187$\,\pm\,$31	 & ~~~~ 45.46$\,\pm\,$1.46 \ \\
 \ 020127	 & ~~~~ 1.9	 & ~~~~ 0.38$\,\pm\,$0.01	 & ~~~~ 290$\,\pm\,$100	 & ~~~~ 46.41$\,\pm\,$1.61 \ \\
 \ 080319C	 & ~~~~ 1.95	 & ~~~~ 1.50$\,\pm\,$0.30	 & ~~~~ 906$\,\pm\,$272	 & ~~~~ 47.14$\,\pm\,$1.57 \ \\
 \ 081008	 & ~~~~ 1.9685	 & ~~~~ 0.96$\,\pm\,$0.09	 & ~~~~ 261$\,\pm\,$52	 & ~~~~ 45.22$\,\pm\,$1.51 \ \\
 \ 030226	 & ~~~~ 1.98	 & ~~~~ 1.30$\,\pm\,$0.10	 & ~~~~ 289$\,\pm\,$66	 & ~~~~ 45.09$\,\pm\,$1.53 \ \\
 \ 000926	 & ~~~~ 2.07	 & ~~~~ 2.60$\,\pm\,$0.60	 & ~~~~ 310$\,\pm\,$20	 & ~~~~ 44.51$\,\pm\,$1.47 \ \\
 \ 090926	 & ~~~~ 2.1062	 & ~~~~ 15.08$\,\pm\,$0.77	 & ~~~~ 974$\,\pm\,$50	 & ~~~~ 44.83$\,\pm\,$1.46 \ \\
 \ 011211	 & ~~~~ 2.14	 & ~~~~ 0.50$\,\pm\,$0.06	 & ~~~~ 186$\,\pm\,$24	 & ~~~~ 45.33$\,\pm\,$1.48 \ \\
 \ 071020	 & ~~~~ 2.145	 & ~~~~ 0.87$\,\pm\,$0.40	 & ~~~~ 1013$\,\pm\,$160	 & ~~~~ 48.015$\,\pm\,$1.49 \ \\
 \ 050922C	 & ~~~~ 2.198	 & ~~~~ 0.47$\,\pm\,$0.16	 & ~~~~ 415$\,\pm\,$111	 & ~~~~ 46.97$\,\pm\,$1.55 \ \\
 \ 060124	 & ~~~~ 2.296	 & ~~~~ 3.40$\,\pm\,$0.50	 & ~~~~ 784$\,\pm\,$285	 & ~~~~ 46.09$\,\pm\,$1.62 \ \\
 \ 021004	 & ~~~~ 2.3	 & ~~~~ 0.27$\,\pm\,$0.04	 & ~~~~ 266$\,\pm\,$117	 & ~~~~ 46.75$\,\pm\,$1.69 \ \\
 \ 051109A	 & ~~~~ 2.346	 & ~~~~ 0.51$\,\pm\,$0.05	 & ~~~~ 539$\,\pm\,$200	 & ~~~~ 47.44$\,\pm\,$1.63 \ \\
 \ 060908	 & ~~~~ 2.43	 & ~~~~ 0.73$\,\pm\,$0.07	 & ~~~~ 514$\,\pm\,$102	 & ~~~~ 46.99$\,\pm\,$1.51 \ \\
 \ 080413	 & ~~~~ 2.433	 & ~~~~ 0.56$\,\pm\,$0.14	 & ~~~~ 584$\,\pm\,$180	 & ~~~~ 47.52$\,\pm\,$1.58 \ \\
 \ 090812	 & ~~~~ 2.452	 & ~~~~ 3.077$\,\pm\,$0.53	 & ~~~~ 2000$\,\pm\,$700	 & ~~~~ 48.06$\,\pm\,$1.61 \ \\
 \ 081121	 & ~~~~ 2.512	 & ~~~~ 1.71$\,\pm\,$0.33	 & ~~~~ 871$\,\pm\,$123	 & ~~~~ 47.11$\,\pm\,$1.49 \ \\
 \ 081118	 & ~~~~ 2.58	 & ~~~~ 0.27$\,\pm\,$0.057	 & ~~~~ 147$\,\pm\,$14	 & ~~~~ 45.69$\,\pm\,$1.47 \ \\
 \ 080721	 & ~~~~ 2.591	 & ~~~~ 7.86$\,\pm\,$1.37	 & ~~~~ 1741$\,\pm\,$227	 & ~~~~ 46.82$\,\pm\,$1.48 \ \\
 \ 050820	 & ~~~~ 2.612	 & ~~~~ 6.40$\,\pm\,$0.50	 & ~~~~ 1325$\,\pm\,$277	 & ~~~~ 46.52$\,\pm\,$1.52 \ \\
 \ 030429	 & ~~~~ 2.65	 & ~~~~ 0.14$\,\pm\,$0.02	 & ~~~~ 128$\,\pm\,$26	 & ~~~~ 46.16$\,\pm\,$1.51 \ \\
 \ 080603B	 & ~~~~ 2.69	 & ~~~~ 0.64$\,\pm\,$0.058	 & ~~~~ 376$\,\pm\,$100	 & ~~~~ 46.60$\,\pm\,$1.55 \ \\
 \ 091029	 & ~~~~ 2.752	 & ~~~~ 0.47$\,\pm\,$0.044	 & ~~~~ 230$\,\pm\,$66	 & ~~~~ 46.00$\,\pm\,$1.56 \ \\
 \ 081222	 & ~~~~ 2.77	 & ~~~~ 1.67$\,\pm\,$0.17	 & ~~~~ 505$\,\pm\,$34	 & ~~~~ 46.16$\,\pm\,$1.47 \ \\
 \ 050603	 & ~~~~ 2.821	 & ~~~~ 3.50$\,\pm\,$0.20	 & ~~~~ 1333$\,\pm\,$107	 & ~~~~ 47.25$\,\pm\,$1.47 \ \\
 \ 050401	 & ~~~~ 2.9	 & ~~~~ 1.90$\,\pm\,$0.40	 & ~~~~ 467$\,\pm\,$110	 & ~~~~ 45.90$\,\pm\,$1.53 \ \\
 \ 090715B	 & ~~~~ 3	 & ~~~~ 1.09$\,\pm\,$0.17	 & ~~~~ 536$\,\pm\,$172	 & ~~~~ 46.80$\,\pm\,$1.59 \ \\
 \ 080607	 & ~~~~ 3.036	 & ~~~~ 8.96$\,\pm\,$0.48	 & ~~~~ 1691$\,\pm\,$226	 & ~~~~ 46.75$\,\pm\,$1.48 \ \\
 \ 081028	 & ~~~~ 3.038	 & ~~~~ 0.81$\,\pm\,$0.095	 & ~~~~ 234$\,\pm\,$93	 & ~~~~ 45.53$\,\pm\,$1.65 \ \\
 \ 020124	 & ~~~~ 3.2	 & ~~~~ 1.20$\,\pm\,$0.10	 & ~~~~ 448$\,\pm\,$148	 & ~~~~ 46.40$\,\pm\,$1.59 \ \\
 \ 060526	 & ~~~~ 3.21	 & ~~~~ 0.12$\,\pm\,$0.06	 & ~~~~ 105$\,\pm\,$21	 & ~~~~ 46.095$\,\pm\,$1.51 \ \\
 \ 080810	 & ~~~~ 3.35	 & ~~~~ 1.82$\,\pm\,$0.20	 & ~~~~ 1470$\,\pm\,$180	 & ~~~~ 48.28$\,\pm\,$1.48 \ \\
 \ 030323	 & ~~~~ 3.37	 & ~~~~ 0.12$\,\pm\,$0.04	 & ~~~~ 270$\,\pm\,$113	 & ~~~~ 47.96$\,\pm\,$1.67 \ \\
 \ 971214	 & ~~~~ 3.42	 & ~~~~ 0.87$\,\pm\,$0.11	 & ~~~~ 685$\,\pm\,$133	 & ~~~~ 47.63$\,\pm\,$1.51 \ \\
 \ 060707	 & ~~~~ 3.425	 & ~~~~ 0.23$\,\pm\,$0.04	 & ~~~~ 279$\,\pm\,$28	 & ~~~~ 47.33$\,\pm\,$1.47 \ \\
 \ 060115	 & ~~~~ 3.53	 & ~~~~ 0.25$\,\pm\,$0.04	 & ~~~~ 285$\,\pm\,$34	 & ~~~~ 47.31$\,\pm\,$1.48 \ \\
 \ 090323	 & ~~~~ 3.57	 & ~~~~ 14.98$\,\pm\,$1.83	 & ~~~~ 1901$\,\pm\,$343	 & ~~~~ 46.55$\,\pm\,$1.50 \ \\
 \ 060206	 & ~~~~ 4.048	 & ~~~~ 0.14$\,\pm\,$0.03	 & ~~~~ 394$\,\pm\,$46	 & ~~~~ 48.68$\,\pm\,$1.48 \ \\
 \ 090516	 & ~~~~ 4.109	 & ~~~~ 1.96$\,\pm\,$0.38	 & ~~~~ 971$\,\pm\,$390	 & ~~~~ 47.58$\,\pm\,$1.65 \ \\
 \ 080916C	 & ~~~~ 4.35	 & ~~~~ 10.13$\,\pm\,$2.13	 & ~~~~ 2646$\,\pm\,$566	 & ~~~~ 47.79$\,\pm\,$1.52 \ \\
 \ 000131	 & ~~~~ 4.5	 & ~~~~ 4.70$\,\pm\,$0.80	 & ~~~~ 987$\,\pm\,$416	 & ~~~~ 46.74$\,\pm\,$1.67 \ \\
 \ 060927	 & ~~~~ 5.6	 & ~~~~ 0.27$\,\pm\,$0.04	 & ~~~~ 475$\,\pm\,$47	 & ~~~~ 48.62$\,\pm\,$1.47 \ \\
 \ 050904	 & ~~~~ 6.29	 & ~~~~ 2$\,\pm\,$0.20	 & ~~~~ 3178$\,\pm\,$1094	 & ~~~~ 50.24$\,\pm\,$1.61 \ \\
 \ 080913	 & ~~~~ 6.695	 & ~~~~ 0.12$\,\pm\,$0.035	 & ~~~~ 710$\,\pm\,$350	 & ~~~~ 50.45$\,\pm\,$1.74 \ \\
 \ 090423	 & ~~~~ 8.1	 & ~~~~ 0.12$\,\pm\,$0.032	 & ~~~~ 491$\,\pm\,$200	 & ~~~~ 49.95$\,\pm\,$1.66 \ \\[1mm]
 \hline\hline
 \end{tabular}
 \end{center}
 \caption{\label{tab2} The numerical data of 59 calibrated
 GRBs at $z>1.4$. The first 4 columns are taken
 from~\cite{r16,r18,r19}, whereas the last column is derived
 by using the calibrated Amati relation. These 59 calibrated
 GRBs are called Hymnium sample, and can be used to constrain
 cosmological models without the circularity problem.}
 \end{table}
 \endgroup



\section{Observational constraints on cosmological models}\label{sec4}


\subsection{Parameterized models}\label{sec4a}

\subsubsection{$\Lambda$CDM model}\label{sec4a1}

\begin{center}
 \begin{figure}[tbhp]
 \centering
 \includegraphics[width=1.0\textwidth]{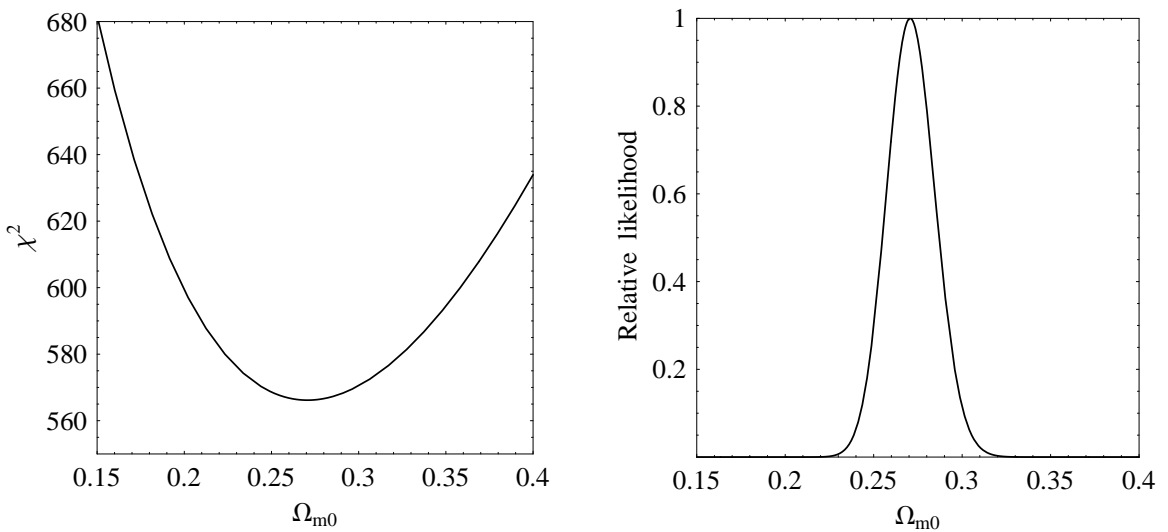}
 \caption{\label{fig3}
 The $\chi^2$ and likelihood ${\cal L}\propto e^{-\chi^2/2}$ as
 functions of $\Omega_{m0}$ for the $\Lambda$CDM model.}
 \end{figure}
 \end{center}

As is well known, for the $\Lambda$CDM model,
 \be{eq17}
 E(z)=\sqrt{\Omega_{m0}(1+z)^3+(1-\Omega_{m0})}\,.
 \ee
 It is easy to obtain the total $\chi^2$ as a function of the
 single model parameter $\Omega_{m0}$ for the $\Lambda$CDM
 model. We present the corresponding $\chi^2$ and likelihood
 ${\cal L}\propto e^{-\chi^2/2}$ in Fig.~\ref{fig3}. The
 best fit has $\chi^2_{min}=566.173$,
 \\[0.7mm] 
 whereas the best-fit parameter is
 $\Omega_{m0}=0.2706^{+0.0138}_{-0.0134}$ (with $1\sigma$
 uncertainty) $^{+0.0282}_{-0.0263}$ (with $2\sigma$ uncertainty).

\vspace{-3mm} 


\vspace{-11mm} 

\subsubsection{XCDM model}\label{sec4a2}

\begin{center}
 \begin{figure}[tbhp]
 \centering
 \includegraphics[width=0.53\textwidth]{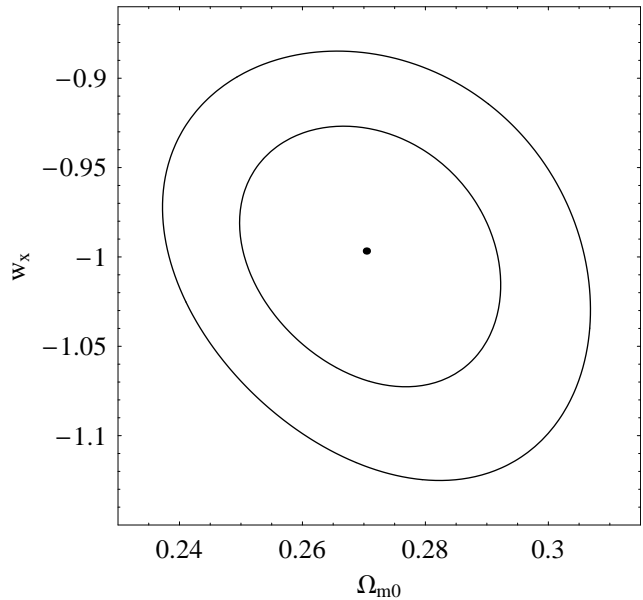}
 \caption{\label{fig4}
 The $68\%$ and $95\%$ confidence level contours in the
 $\Omega_{m0}-w_{_X}$ parameter space for the XCDM model.
 The best-fit parameters are also indicated by a black
 solid point.}
 \end{figure}
 \end{center}

Next, we consider the XCDM model. It is also well known that
 in the spatially flat universe which contains pressureless
 matter and dark energy whose equation-of-state parameter (EoS)
 is a constant $w_{_X}$, the corresponding $E(z)$ is given by
 \be{eq18}
 E(z)=\sqrt{\Omega_{m0}(1+z)^3+(1-\Omega_{m0})(1+z)^{3(1+w_{_X})}}\,.
 \ee
 By minimizing the corresponding total $\chi^2$ in Eq.~(\ref{eq16}),
 we find the best-fit parameters $\Omega_{m0}=0.2704$ and
 $w_{_X}=-0.9967$, while $\chi^2_{min}=566.168$. In Fig.~\ref{fig4},
 we present the corresponding $68\%$ and $95\%$ confidence
 level contours in the $\Omega_{m0}-w_{_X}$ parameter space for
 the XCDM model.

\subsubsection{CPL model}\label{sec4a3}

\begin{center}
 \begin{figure}[tbph]
 \centering
 \includegraphics[width=0.48\textwidth]{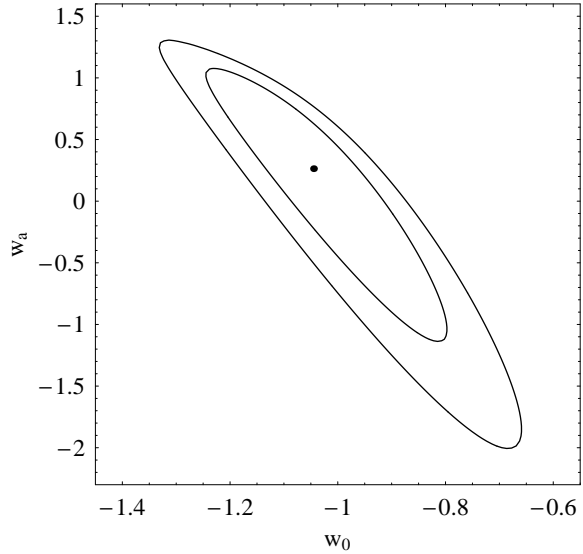}
 \caption{\label{fig5}
 The $68\%$ and $95\%$ confidence level contours in the
 $w_0-w_a$ plane for the CPL model. The best-fit parameters
 are also indicated by a black solid point.}
 \end{figure}
 \end{center}

\begin{center}
 \begin{figure}[htbp]
 \centering
 \vspace{3mm} 
 \includegraphics[width=1.0\textwidth]{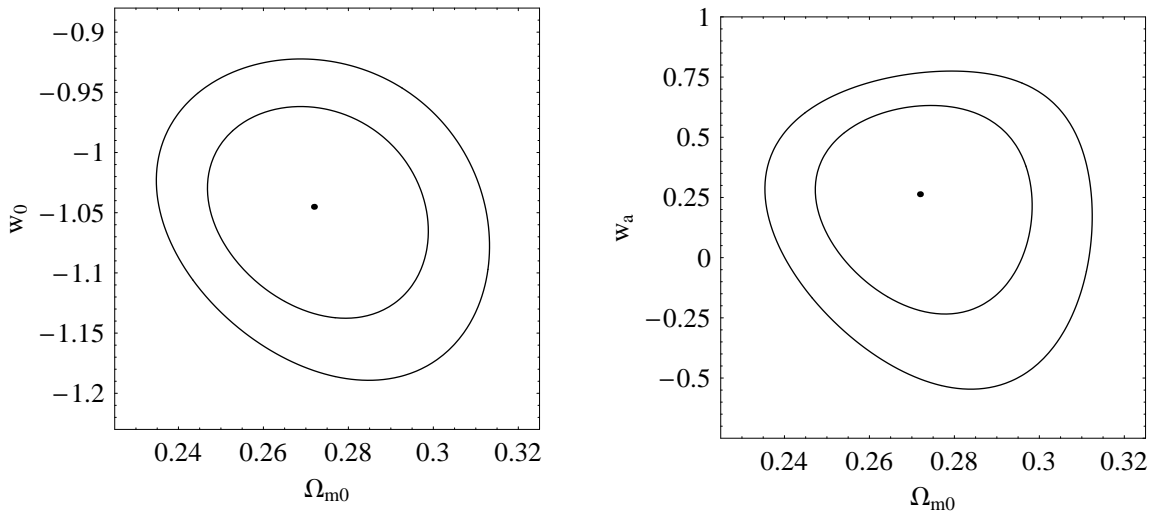}
 \caption{\label{fig6}
 The same as in Fig.~\ref{fig5}, except for the $\Omega_{m0}-w_0$
 plane and the $\Omega_{m0}-w_a$ plane.}
 \end{figure}
 \end{center}

Here, we consider the familiar Chevallier-Polarski-Linder
 (CPL) model~\cite{r35}, in which the EoS of dark energy is
 parameterized as
 \be{eq19}
 w_{de}=w_0+w_a(1-a)=w_0+w_a\frac{z}{1+z}\,,
 \ee
 where $w_0$ and $w_a$ are constants. As is well known, the
 corresponding $E(z)$ is given by~\cite{r34,r36,r37}
 \be{eq20}
 E(z)=\left[\Omega_{m0}(1+z)^3
 +\left(1-\Omega_{m0}\right)(1+z)^{3(1+w_0+w_a)}
 \exp\left(-\frac{3w_a z}{1+z}\right)\right]^{1/2}.
 \ee
 There are 3 independent parameters in this model. By minimizing the
 corresponding total $\chi^2$ in Eq.~(\ref{eq16}), we find the
 best-fit parameters $\Omega_{m0}=0.2719$, $w_0=-1.0451$
 and $w_a=0.2635$, while $\chi^2_{min}=566.007$. In Fig.~\ref{fig5},
 we present the corresponding $68\%$ and $95\%$ confidence level
 contours in the $w_0-w_a$ plane for the CPL model. Also,
 the $68\%$ and $95\%$ confidence level contours in the
 $\Omega_{m0}-w_0$ plane and the $\Omega_{m0}-w_a$ plane for
 the CPL model are shown in Fig.~\ref{fig6}.


\vspace{-11mm} 


\subsection{DGP model}\label{sec4b}

\begin{center}
 \begin{figure}[tbhp]
 \centering
 \includegraphics[width=1.0\textwidth]{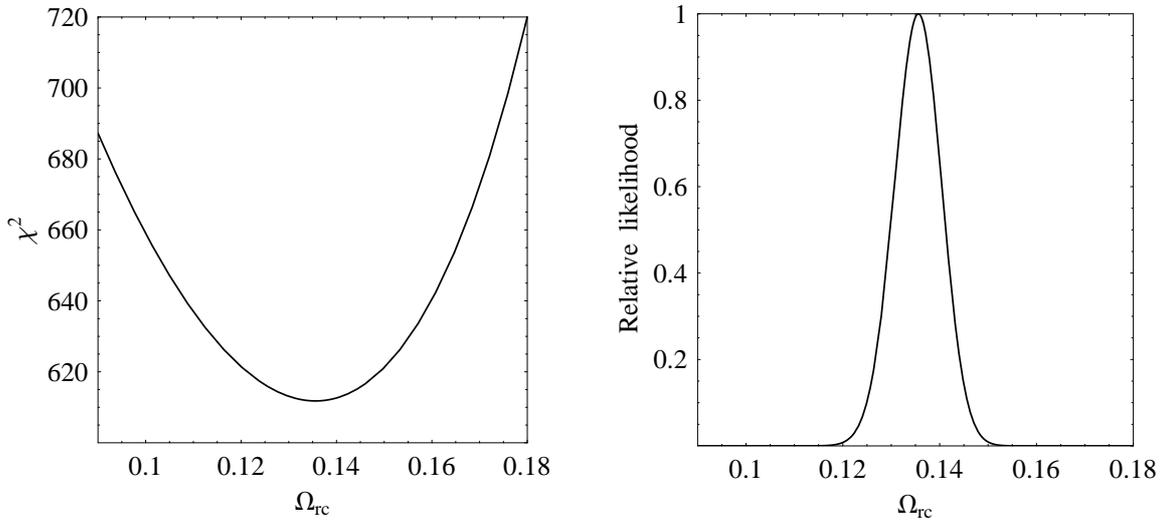}
 \caption{\label{fig7}
 The $\chi^2$ and likelihood ${\cal L}\propto e^{-\chi^2/2}$ as
 functions of $\Omega_{rc}$ for the DGP model.}
 \end{figure}
 \end{center}

One of the simplest modified gravity models is the so-called
 Dvali-Gabadadze-Porrati~(DGP) braneworld model~\cite{r38,r39},
 which entails altering the Einstein-Hilbert action by a term
 arising from large extra dimensions. For a list of references
 on the DGP model, see e.g.~\cite{r40,r41} and references
 therein. As is well known, for the spatially flat DGP model
 (here we only consider the self-accelerating branch), $E(z)$
 is given by~\cite{r39,r40,r41}
 \be{eq21}
 E(z)=\sqrt{\Omega_{m0}(1+z)^3+\Omega_{rc}}+
 \sqrt{\Omega_{rc}}\,,
 \ee
 where $\Omega_{rc}$ is a constant. It is easy to see that
 $E(z=0)=1$ requires
 \be{eq22}
 \Omega_{m0}=1-2\sqrt{\Omega_{rc}}\,.
 \ee
 Therefore, the DGP model has only one independent model parameter
 $\Omega_{rc}$. Notice that $0\leq\Omega_{rc}\leq 1/4$ is required
 by $0\leq\Omega_{m0}\leq 1$. It is easy to obtain the total
 $\chi^2$ as a function of the single model parameter
 $\Omega_{rc}\,$. In Fig.~\ref{fig7}, we plot the corresponding
 $\chi^2$ and likelihood ${\cal L}\propto e^{-\chi^2/2}$. The
 best fit has $\chi^2_{min}=611.794$,
 \\[0.7mm] 
 whereas the best-fit parameter is
 $\Omega_{rc}=0.1356^{+0.0048}_{-0.0049}$ (with $1\sigma$
 uncertainty) $^{+0.0095}_{-0.0099}$ (with $2\sigma$ uncertainty).


\vspace{-10mm} 


\vspace{-10mm} 


\subsection{New agegraphic dark energy model}\label{sec4c}

In~\cite{r42,r43}, the so-called new agegraphic dark energy
 (NADE) model has been proposed recently, based on the
 K\'{a}rolyh\'{a}zy uncertainty relation which arises from
 quantum mechanics together with general relativity. In this
 model, the energy density of NADE is given by~\cite{r42,r43}
 \be{eq23}
 \rho_q=\frac{3n^2m_p^2}{\eta^2}\,,
 \ee
 where $m_p$ is the reduced Planck mass; $n$ is a constant of
 order unity; $\eta$ is the conformal time
 \be{eq24}
 \eta\equiv\int\frac{dt}{a}=\int\frac{da}{a^2H}\,,
 \ee
 in which $a=(1+z)^{-1}$ is the scale factor. Obviously,
 $\dot{\eta}=1/a$, where a dot denotes the derivative with
 respect to cosmic time $t$. The corresponding fractional
 energy density of NADE reads
 \be{eq25}
 \Omega_q\equiv\frac{\rho_q}{3m_p^2H^2}=\frac{n^2}{H^2\eta^2}\,.
 \ee
 From the Friedmann equation
 $H^2=\left(\rho_m+\rho_q\right)/\left(3m_p^2\right)$,
 the energy conservation equation $\dot{\rho}_m+3H\rho_m=0$,
 and Eqs.~(\ref{eq23})---(\ref{eq25}), we find that the
 equation of motion for $\Omega_q$ is given by~\cite{r42,r43}
 \be{eq26}
 \frac{d\Omega_q}{dz}=-\Omega_q\left(1-\Omega_q\right)
 \left[3(1+z)^{-1}-\frac{2}{n}\sqrt{\Omega_q}\right].
 \ee
 From the energy conservation
 equation $\dot{\rho}_q+3H(\rho_q+p_q)=0$, and
 Eqs.~(\ref{eq23})---(\ref{eq25}), it is easy to find that
 the EoS of NADE is given by~\cite{r42,r43}
 \be{eq27}
 w_q\equiv\frac{p_q}{\rho_q}=
 -1+\frac{2}{3n}\frac{\sqrt{\Omega_q}}{a}\,.
 \ee
 When $a\to\infty$, $\Omega_q\to 1$, thus $w_q\to -1$ in the
 late time. When $a\to 0$, $\Omega_q\to 0$, so $0/0$ appears in
 $w_q$ and hence we cannot directly obtain $w_q$
 from Eq.~(\ref{eq27}). Let us consider the matter-dominated
 epoch, in which $H^2\propto\rho_m\propto a^{-3}$. Thus,
 $a^{1/2}da\propto dt=ad\eta$. Then, we have $\eta\propto a^{1/2}$.
 From Eq.~(\ref{eq23}), $\rho_q\propto a^{-1}$.  From the energy
 conservation equation $\dot{\rho}_q+3H\rho_q(1+w_q)=0$, we
 obtain $w_q=-2/3$ in the matter-dominated epoch. Since
 $\rho_m\propto a^{-3}$ and $\rho_q\propto a^{-1}$, it is
 expected that $\Omega_q\propto a^2$. Comparing $w_q=-2/3$ with
 Eq.~(\ref{eq27}), we find that $\Omega_q=n^2a^2/4$ in the
 matter-dominated epoch as expected. For $a\ll 1$, provided
 that $n$ is of order unity, $\Omega_q\ll 1$ naturally follows.
 There are many interesting features in the NADE model and we
 refer to the original papers~\cite{r42,r43} for more details.

At first glance, one might consider that NADE is
 a two-parameter model. However, as shown in~\cite{r42}, NADE
 is a {\em single-parameter} model in practice, thanks to its
 special analytic feature $\Omega_q=n^2a^2/4=n^2(1+z)^{-2}/4$
 in the matter-dominated epoch, as mentioned above. If $n$ is
 given, we can obtain $\Omega_q(z)$ from Eq.~(\ref{eq26}) with
 the initial condition $\Omega_q(z_{ini})=n^2(1+z_{ini})^{-2}/4$ at
 any $z_{ini}$ which is deep enough into the matter-dominated
 epoch (we choose $z_{ini}=2000$ as in~\cite{r42}), instead of
 $\Omega_q(z=0)=1-\Omega_{m0}$ at $z=0$. Then, all other
 physical quantities, such as $\Omega_m(z)=1-\Omega_q(z)$
 and $w_q(z)$ in Eq.~(\ref{eq27}), can be obtained correspondingly.
 So, $\Omega_{m0}=\Omega_m(z=0)$, $\Omega_{q0}=\Omega_q(z=0)$ and
 $w_{q0}=w_q(z=0)$ are {\em not} independent model parameters. The
 only free model parameter is $n$. Therefore, the NADE model is a
 {\em single-parameter} model in practice.

\begin{center}
 \begin{figure}[tbhp]
 \centering
 \includegraphics[width=1.0\textwidth]{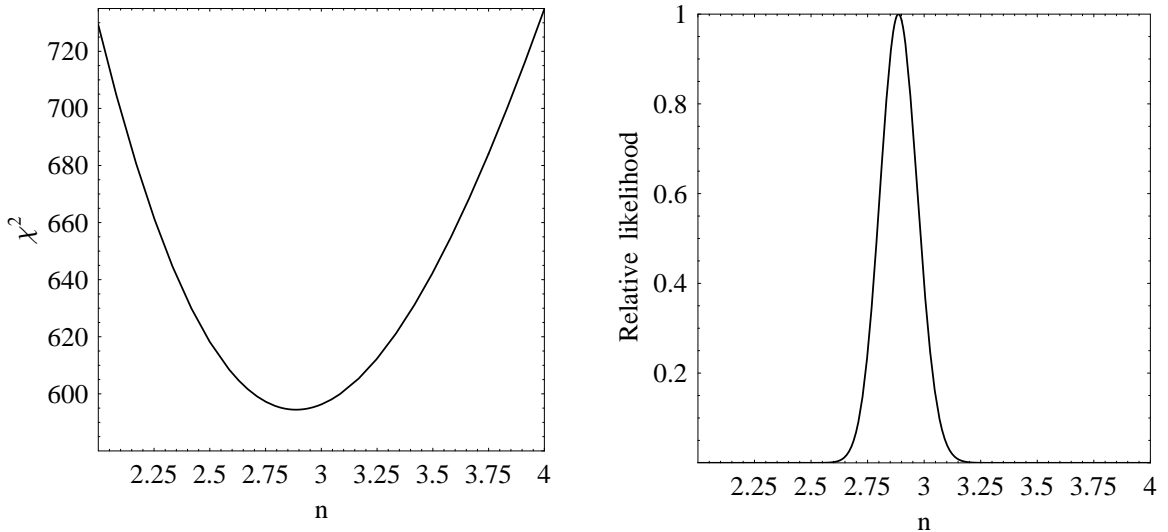}
 \caption{\label{fig8}
 The $\chi^2$ and likelihood ${\cal L}\propto e^{-\chi^2/2}$ as
 functions of $n$ for the NADE model.}
 \end{figure}
 \end{center}

From the Friedmann equation
 $H^2=\left(\rho_m+\rho_q\right)/\left(3m_p^2\right)$, we have
 \be{eq28}
 E(z)=\left[\frac{\Omega_{m0}(1+z)^3}{1-\Omega_q(z)}\right]^{1/2}.
 \ee
 If the single model parameter $n$ is given, we can obtain
 $\Omega_q(z)$ from Eq.~(\ref{eq26}). Thus, we get
 $\Omega_{m0}=1-\Omega_q(z=0)$. So, $E(z)$ is at hand.
 Therefore, we can find the corresponding total $\chi^2$ in
 Eq.~(\ref{eq16}). In Fig.~\ref{fig8}, we plot the total
 $\chi^2$ and likelihood ${\cal L}\propto e^{-\chi^2/2}$ as
 functions of $n$. The best fit has $\chi^2_{min}=594.449$,
 \\[0.7mm] 
 whereas the best-fit parameter is $n=2.8865^{+0.0832}_{-0.0818}$
 (with $1\sigma$ uncertainty) $^{+0.1680}_{-0.1622}$ (with $2\sigma$
 uncertainty).


\vspace{-11mm} 


\subsection{Holographic dark energy model}\label{sec4d}

\begin{center}
 \begin{figure}[tbhp]
 \centering
 \includegraphics[width=0.5\textwidth]{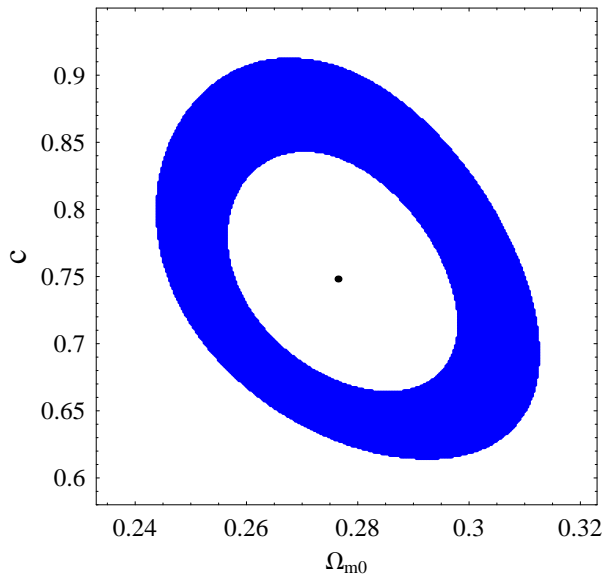}
 \caption{\label{fig9}
 The $68\%$ and $95\%$ confidence level contours in the
 $\Omega_{m0}-c$ parameter space for the HDE model. The
 best-fit parameters are also indicated by a black solid point.}
 \end{figure}
 \end{center}

The so-called holographic dark energy (HDE) has been studied
 extensively in the literature. It is proposed from the holographic
 principle~\cite{r44,r45} in the string theory. For a quantum
 gravity system, the local quantum field cannot contain too
 many degrees of freedom, otherwise the formation of black
 hole is inevitable and then the quantum field theory breaks
 down. In the thermodynamics of the black hole~\cite{r46,r47},
 there is a maximum entropy in a box of size $L$, namely the
 so-called Bekenstein entropy bound $S_{BH}$, which scales as
 the area of the box $\sim L^2$, rather than the volume
 $\sim L^3$. To avoid the breakdown of the local quantum field
 theory, Cohen {\it et al.}~\cite{r48} proposed a more
 restrictive bound, i.e., the energy bound. If $\rho_\Lambda$
 is the quantum zero-point energy density caused by a short
 distance cut-off, the total energy in a box of size $L$ cannot
 exceed the mass of a black hole of the same size~\cite{r48},
 namely $L^3\rho_\Lambda\,\lsim\, Lm_p^2$. The largest IR
 cut-off $L$ is the one saturating the inequality. Thus,
 \be{eq29}
 \rho_\Lambda=3c^2\,m_p^2\,L^{-2},
 \ee
 where the numerical constant $3c^2$ is introduced for convenience.
 For the HDE model proposed in~\cite{r49}, the cut-off $L$ has been
 chosen to be the future event horizon $R_h$, which is given by
 \be{eq30}
 R_h=a\int_t^\infty\frac{d\tilde{t}}{a}
 =a\int_a^\infty\frac{d\tilde{a}}{H\tilde{a}^2}\,.
 \ee
 From Eqs.~(\ref{eq29}), (\ref{eq30}), and the energy conservation
 equation $\dot{\rho}_\Lambda+3H\rho_\Lambda(1+w_\Lambda)=0$, it is
 easy to find that (see e.g.~\cite{r49,r50,r51,r52,r53})
 \be{eq31}
 \frac{d\Omega_\Lambda}{dz}=-(1+z)^{-1}\Omega_\Lambda
 (1-\Omega_\Lambda)\left(1+\frac{2}{c}\sqrt{\Omega_\Lambda}\right),
 \ee
 where $\Omega_\Lambda$ is the fractional energy density of HDE.
 From the Friedmann equation
 $H^2=\left(\rho_m+\rho_\Lambda\right)/\left(3m_p^2\right)$, we have
 \be{eq32}
 E(z)=
 \left[\frac{\Omega_{m0}(1+z)^3}{1-\Omega_\Lambda(z)}\right]^{1/2}.
 \ee
 There are 2 independent model parameters, namely $\Omega_{m0}$
 and $c\,$. One can obtain $\Omega_\Lambda(z)$ by solving the
 differential equation~(\ref{eq31}) with the initial condition
 $\Omega_\Lambda(z=0)=1-\Omega_{m0}$. Substituting
 $\Omega_\Lambda(z)$ into Eq.~(\ref{eq32}), we can find the
 corresponding $E(z)$ and then the total $\chi^2$
 in Eq.~(\ref{eq16}). By minimizing the total $\chi^2$, we find the
 best-fit parameters $\Omega_{m0}=0.2764$ and $c=0.7482$,
 while $\chi^2_{min}=566.215$. In Fig.~\ref{fig9},
 we present the corresponding $68\%$ and $95\%$ confidence
 level contours in the $\Omega_{m0}-c$ parameter space for
 the HDE model.


\vspace{-11mm} 


\subsection{Ricci dark energy model}\label{sec4e}

\begin{center}
 \begin{figure}[tbhp]
 \centering
 \includegraphics[width=0.5\textwidth]{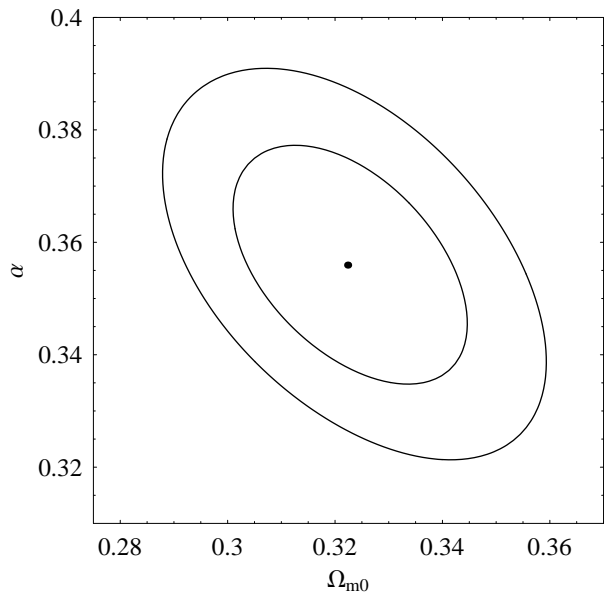}
 \caption{\label{fig10}
 The $68\%$ and $95\%$ confidence level contours in the
 $\Omega_{m0}-\alpha$ parameter space for the RDE model.
 The best-fit parameters are also indicated by a black
 solid point.}
 \end{figure}
 \end{center}

The so-called Ricci dark energy (RDE) model was proposed
 in~\cite{r54}, which can be regarded as a variant of HDE
 mentioned above, while its corresponding cut-off $L$ in
 Eq.~(\ref{eq29}) is chosen to be proportional to the Ricci
 scalar curvature radius. In~\cite{r54}, there is no physical
 motivation to this proposal for $L$ in fact. Recently,
 in~\cite{r55} it is found that the Jeans length $R_{\rm CC}$
 which is determined by $R_{\rm CC}^{-2}=\dot{H}+2H^2$ gives
 the causal connection scale of perturbations in the flat
 universe. Since the Ricci scalar is also proportional to
 $\dot{H}+2H^2$ in the flat universe, the physical motivation
 for RDE has been found in~\cite{r55} actually. In the RDE
 model, the corresponding $\rho_\Lambda$ is given
 by~\cite{r54}
 \be{eq33}
 \rho_\Lambda=3\alpha m_p^2 \left(\dot{H}+2H^2\right),
 \ee
 where $\alpha$ is a positive constant (when $L$ is chosen
 to be $R_{\rm CC}$ in Eq.~(\ref{eq29}), one can see that
 $\alpha=c^2$ in fact). Substituting Eq.~(\ref{eq33}) into
 Friedmann equation, it is easy to find that~\cite{r54,r53}
 \be{eq34}
 E(z)=\left[\,\frac{2\Omega_{m0}}{\,2-\alpha}\,(1+z)^3+
 \left(1-\frac{2\Omega_{m0}}{\,2-\alpha}\right)
 (1+z)^{4-2/\alpha}\right]^{1/2}.
 \ee
 There are 2 independent model parameters, namely $\Omega_{m0}$
 and $\alpha\,$. By minimizing the corresponding total $\chi^2$
 in Eq.~(\ref{eq16}), we find the best-fit parameters
 $\Omega_{m0}=0.3223$ and $\alpha=0.3559$, while
 $\chi^2_{min}=589.026$. In Fig.~\ref{fig10}, we present the
 corresponding $68\%$ and $95\%$ confidence level contours in
 the $\Omega_{m0}-\alpha$ parameter space for the RDE model.


\vspace{-10mm} 


\section{Comparison of models}\label{sec5}

In the previous section, we have obtained the constraints on
 7 cosmological models with the latest observational data,
 namely, the combination of 557 Union2 SNIa dataset~\cite{r20},
 59 calibrated Hymnium GRBs dataset (obtained in this work),
 the shift parameter $R$ from the WMAP7 data~\cite{r21}, and
 the distance parameter $A$ of the measurement of the BAO peak
 in the distribution of SDSS luminous red galaxies~\cite{r22,r23}.
 In Table~\ref{tab3}, we summarize the results for these 7 models.
 Here, we would like to briefly consider the comparison of
 these models. Naively, one can compare them with their
 $\chi^2_{min}$, i.e., the best model has the smallest
 $\chi^2_{min}$ (vice versa). However, as is well known, in
 general $\chi^2_{min}$ decreases when the number of free model
 parameters increases. So, $\chi^2_{min}$ is not a good
 criterion for model comparison. Instead, a conventional
 criterion for model comparison in the literature
 is $\chi^2_{min}/dof$, in which the degree of freedom
 $dof=N-k$, whereas $N$ and $k$ are the number of data points
 and the number of free model parameters, respectively. We
 present the $\chi^2_{min}/dof$ for all the 7 models in
 Table~\ref{tab3}. On the other hand, there are
 other criterions for model comparison in the literature. The
 most sophisticated criterion is the Bayesian evidence (see
 e.g.~\cite{r56} and references therein). However, the
 computation of Bayesian evidence usually consumes a large
 amount of time and power. As an alternative, one can consider
 some approximations of Bayesian evidence, such as the
 so-called Bayesian Information Criterion (BIC) and Akaike
 Information Criterion (AIC). The BIC is defined by~\cite{r57}
 \be{eq35}
 {\rm BIC}=-2\ln{\cal L}_{max}+k\ln N\,,
 \ee
 where ${\cal L}_{max}$ is the maximum likelihood. In the
 Gaussian cases, $\chi^2_{min}=-2\ln{\cal L}_{max}$. So, the
 difference in BIC between two models is given by
 $\Delta{\rm BIC}=\Delta\chi^2_{min}+\Delta k \ln N$. The AIC
 is defined by~\cite{r58}
 \be{eq36}
 {\rm AIC}=-2\ln{\cal L}_{max}+2k\,.
 \ee
 The difference in AIC between two models is given by
 $\Delta{\rm AIC}=\Delta\chi^2_{min}+2\Delta k$. We refer to
 e.g.~\cite{r52,r59} for some relevant works with BIC and AIC.
 In Table~\ref{tab3}, we also present the $\Delta$BIC and
 $\Delta$AIC of all the 7 models considered in this work.
 Notice that $\Lambda$CDM has been chosen to be the fiducial
 model when we calculate $\Delta$BIC and $\Delta$AIC.
 From Table~\ref{tab3}, it is easy to see that the rank of
 models is coincident in all the 3 criterions
 ($\chi^2_{min}/dof$, BIC and AIC). The $\Lambda$CDM model
 is still the best one, whereas DGP model is the worst one.
 The detailed rank of all the 7 models is also given
 in Table~\ref{tab3}.


 \begin{table}[tb]
 \begin{center}
 \begin{tabular}{llllllll}\hline\hline
 ~~Model~~~~~~~~ & $\Lambda$CDM~~~~ & XCDM~~~~ & CPL~~~~~~
 & DGP~~~~~~ & NADE~~~~ & HDE~~~~~~ & RDE~~~~~~\\ \hline
 ~~$\chi^2_{min}$ & 566.173 & 566.168 & 566.007 & 611.794
 & 594.449 & 566.215 & 589.026\\
 ~~$k$ & 1 & 2 & 3 & 1 & 1 & 2 & 2\\
 ~~$\chi^2_{min}/dof$ & 0.918 & 0.919 & 0.920 & 0.992 & 0.963
 & 0.919 & 0.956\\
 ~~$\Delta$BIC & 0 & 6.421 & 12.687 & 45.621 & 28.276 & 6.468
 & 29.280\\
 ~~$\Delta$AIC & 0 & 1.995 & 3.834 & 45.621 & 28.276 & 2.042
 & 24.853\\
 ~~Rank & 1 & $2\sim 3$ & 4 & 7 & $5\sim 6$ & $2\sim 3$
 & $5\sim 6$\\
 \hline\hline
 \end{tabular}
 \end{center}
 \caption{\label{tab3} Summarizing all the 7 models considered
 in this work.}
 \end{table}



\section{Conclusions}\label{sec6}

In the present work, by the help of the newly released Union2
 compilation which consists of 557 SNIa, we calibrated 109
 long GRBs with the well-known Amati relation, using the
 cosmology-independent calibration method proposed
 in~\cite{r12}. We have obtained 59 calibrated high-redshift
 GRBs which can be used to constrain cosmological models
 without the circularity problem (we call them Hymnium GRBs
 sample for convenience). One can directly read off the
 numerical data of these 59 Hymnium GRBs from Table~\ref{tab2}
 of this paper. They are also available upon request to our
 email address. Considering their unique ability to extend our
 vision up to redshift $z=8.1$, we recommend to use these 59
 Hymnium GRBs in the relevant works.

We considered the joint constraints on 7 cosmological models
 from the latest observational data, namely, the combination of
 557 Union2 SNIa dataset~\cite{r20}, 59 calibrated Hymnium GRBs
 dataset (obtained in this work), the shift parameter $R$ from
 the WMAP7 data~\cite{r21}, and the distance parameter $A$ of
 the measurement of the BAO peak in the distribution of SDSS
 luminous red galaxies~\cite{r22,r23}. Note that in general it
 is better to see how the parameters are constrained both with
 and without GRBs. However, in fact the cosmological
 constraints on the same models without GRBs have been
 considered in e.g.~\cite{r52}, although they used slightly
 earlier SNIa and CMB data. So, we do not consider the
 constraints without GRBs in the present work. Comparing with
 the previous results in the literature (e.g.~\cite{r52}), one
 can find that the constraints obtained in this work are
 tighter, thanks to these newly improved observational data.
 We also briefly considered the comparison of these 7 cosmological
 models. The $\Lambda$CDM model is still the best one from the
 perspective of the latest cosmological observations.


\section*{ACKNOWLEDGEMENTS}
We thank the anonymous referee for quite useful comments and
 suggestions, which help us to improve this work. We are
 indebted to Lorenzo~Amati for providing us the data of
 14 unpublished GRBs in private communication and kindly
 allowing us to use them in our relevant works. We are grateful
 to Professors Rong-Gen~Cai and Shuang~Nan~Zhang for helpful
 discussions. We also thank Minzi~Feng, as well as Xiao-Peng~Ma
 and Bo~Tang, for kind help and discussions. This work was
 supported in part by NSFC under Grant No.~10905005, the
 Excellent Young Scholars Research Fund of Beijing Institute
 of Technology, and the Fundamental Research Fund of Beijing
 Institute of Technology.

\renewcommand{\baselinestretch}{1.1}



\begin{thebibliography}{99}
\bibitem{r60}
In fact, up to now, there are more than 109 long GRBs with
 known redshifts and spectra peak energy, see e.g.~\cite{r61}
 and references therein. However, since the numerical data
 of $E_{\rm p,i}$ and $S_{\rm bolo}$ are explicitly given
 in~\cite{r17} (and~\cite{r16,r18,r19}) which are more
 convenient for our computing, we choose to use these 109
 long GRBs instead. One can easily extend our work with a
 larger GRBs sample.
\end{thebibliography}
\end{document}